\documentclass[letterpaper]{article}
\usepackage{aaai17}
\usepackage{times}
\usepackage{helvet}
\usepackage{courier}
\usepackage{multirow}
\usepackage{todonotes}
\usepackage{amsmath, amsfonts}
\usepackage{graphicx}
\usepackage{color}
\usepackage{epstopdf}
\usepackage{wrapfig}

\usepackage{algorithm}
\usepackage[noend]{algpseudocode}

\usepackage{mathtools}
\usepackage{caption}
\usepackage{morefloats}
\usepackage{subfloat}
\usepackage{booktabs}
\usepackage{listings}
\usepackage{parcolumns}
\usepackage{placeins}
\usepackage{natbib}
\usepackage{url}

\newcommand{\blue}[1]{{\bf{\textcolor{blue}{#1}}}}

\begin{document}

\title{Probabilistic Inference of Twitter Users' Age based on What They Follow}

\author{
% Paper-ID 245
Benjamin Paul Chamberlain$^1$\hspace{5mm}  Clive Humby$^2$ \hspace{5mm} Marc Peter Deisenroth$^1$\\
$^1$Department of Computing, Imperial College London, London, UK\\
$^2$Starcount Insights, 2 Riding House Street, London, UK
}
\maketitle

\begin{abstract}

Twitter provides an open and rich source of data for studying human behaviour at scale and is widely used in social and network sciences. However, a major criticism of Twitter data is that demographic information is largely absent. Enhancing Twitter data with user ages would advance our ability to study social network structures, information flows and the spread of contagions. 
% how this is usually solved
Approaches toward age detection of Twitter users typically focus on specific properties of tweets, e.g., linguistic features, which are language dependent.
% what we do
In this paper, we devise a language-independent methodology for determining the age of Twitter users from data that is native to the Twitter ecosystem. The key idea is to use a Bayesian framework to generalise ground-truth age information from a few Twitter users to the entire network based on what/whom they follow.
% punchline
Our approach scales to inferring the age of 700 million Twitter accounts with high accuracy.
%\keywords{social networks, age detection}
\end{abstract}

\section{Introduction}
% social media data
Digital social networks (DSNs) produce data that is of great scientific value. They have allowed researchers to study the flow of information, the structure of society and major political events (e.g., the Arab Spring) quantitatively at scale.

% Intro to Twitter
Owing to its simplicity, size and openness, Twitter is among the most popular DSNs used for scientific research. On the Twitter platform users generate data by \emph{tweeting} a stream of 140 character (or less) messages. To consume content users \emph{follow} each other. Following is a one-way interaction, and for this reason Twitter is regarded as an \emph{interest network}~\citep{Gupta2013}.  By default, Twitter is entirely public, and there are no requirements for users to enter personal information.

% issues with Twitter
The lack of reliable (or usually any) demographic data is a major criticism of the usefulness of Twitter for research purposes. Enriching Twitter accounts with demographic information (e.g., age) would be valuable for scientific, industrial and governmental applications. Explicit examples include opinion polling, product evaluations and market research. 

% Homophily
Our assumes that people who are close in age have similar interests as a result of age-related life events (e.g., education, child birth, marriage, employment, retirement, wealth changes). This is an example of the well-known homophily principle, which states that people with related attributes form similar ties~\citep{McPherson2001}. For age inference in Twitter, we exploit that most Follows\footnote{we use capitalisation to indicate the Twitter specific usage of this word} are indicative of a user's interests. Putting things together, we arrive at our central hypothesis that (a) somebody follows what is interesting to them, (b) their interests are indicative of their age. Hence, we propose to infer somebody's  age based on  what/whom they Follow. We created the artificial @williamockam account shown in Figure~\ref{fig:twitter_profile} to use as a running example to illustrate our method.

% our contribution
The contribution of this paper is the derivation of a probabilistic model that infers any Twitter user's age only based on what/whom they Follow, which is not restricted by national and linguistic boundaries. Our model handles the high levels of noise in the data in a principled way and is massively scalable allowing us to infer the age of 700 million Twitter accounts with high accuracy. In addition we supply a new public dataset for use by researchers interested in the problem of attributing vertices in social networks.

\section{Related Work}
\label{sec:rel}
There is a large body of excellent research on enhancing social data with demographic attributes. 
% learning demographics on Twitter
This includes work on gender~\citep{Burger2011}, political affiliation~\citep{Conover2011, Pennacchiotti2011}, location~\citep{Cheng2010} and ethnicity~\citep{Mislove2011, Chang2010, Pennacchiotti2011}. Also of note is the work of \cite{Fang2015} who focus on modelling the correlations between various demographic attributes.

% AGE DETECTION
% image based age detection
Some of the most exciting recent work on detecting ages from social data has been in the field of computer vision where age is determined from user images~\citep{Fu2010, Guo2008}. However, computer vision methods are difficult to apply to Twitter data: few accounts have profile images and those that do are often inaccurate or of poor quality.

% previous work on age detection on Twitter
Following the seminal work of \cite{Schler2006}, the majority of research on age detection of Twitter users has focused on linguistic models of tweets~\citep{Nguyen2011,Rao2010, AlZamal2012}. Notably, \cite{nguyen2013old} developed a linguistic model for Dutch tweets that allows them to predict the age category (using logistic regression) of Twitter users who have tweeted more than ten times in Dutch. They performed a lexical analysis of Dutch language tweets and obtained ground truth through a labour intensive manual tagging process. The principal features were unigrams, assuming that older people use more positive language, fewer pronouns and longer sentences. They concluded that age prediction works well for young people, but that above the age of 30, language tends to homogenise. 

% age detection with lexical features: social-age problem
In general, lexical approaches suffer from the concept of social age~\citep{DongNguyen142014}. Social age is determined by life stage (married, children, employment etc.) rather than years since birth, and it has a strong affect on writing style. People often adapt their language to mimic the perceived social norm in a group. 
% age detection based on tweets  has some limitations
Additionally, tweet-based methods struggle to make predictions for Twitter users with low tweet counts. In practice, this is a major problem since we calculated that the median number of tweets for the 700m Twitter users in our data set is only 4 (the \emph{tweets} field shown in Fig.~\ref{fig:twitter_profile} is available as account metadata for all accounts). 

% working with user names
The user name has also been considered as a source of demographic information. This was first done by \cite{Liu2013} to detect gender and later by \cite{Oktay2014} to estimate the age of Twitter users from the first name supplied in the free-text \emph{account name} field (eg. William in Figure~\ref{fig:twitter_profile}). In their research, they use US social security data to generate probability distributions of birth years given age. They show that for some names age distributions are quite sharply peaked. 
%They also found that both age and ethnicity affect the patterns of Twitter usage.
% user name field problems
A potential issue with this approach is that methods based on the ``user name'' field rely on knowledge of the user's true first name and their country of birth~\citep{Oktay2014}. In practice, this assumption is problematic since Twitter users often do not use their real names, and their country of birth is generally unknown. 

% using the Facebook like network
Approaches to combine lexical and network features include \cite{AlZamal2012, Pennacchiotti2011}, who show that using the graph structure can improve performance at the expense of scalability.
\cite{Kosinski2013} used Face\-book-Likes to predict a broad range of user attributes mined from 58,466 survey correspondents in the US. Their approach of solely using Facebook Likes as features for learning has the benefit of generalising readily to different locales. \cite{Culotta2015} have applied a similar Follower based approach to Twitter to predict demographic attributes, however their approach of using aggregate distributions of website visitors as ground-truth is restricted to predicting the aggregate age of groups of users.
% how our assumptions/settings differ from the Facebook setting
% Our work is inspired by the generality these approaches as Facebook Likes are closely related to Twitter Follows: Both are indicators of user interest. However, the setting in \citep{Kosinski2013} differs from ours in two main aspects: Firstly, Facebook explicitly contains age information (birthdays). Secondly, our approach generalises from a few labelled samples to the entire population. Since the Twitter users who explicitly provide age information are likely to be a highly biased (younger) sample of the population, we must explicitly account for this bias to make good population predictions.  
Our work is inspired by the generality of the approaches of \cite{Kosinski2013} and \cite{Culotta2015}, however our setting differs in two ways. We use data native to the Twitter ecosystem to generalise from a few examples to make individual predictions for the entire Twitter population. Secondly we do not make the assumption that our sample is an unbiased estimate of the Twitter population and we explicitly account for this bias to make good population predictions. For these reasons it is hard to get ground truth and careful probabilistic modelling is required to infer the age of arbitrary Twitter users.

%%%%%%%%%%%%%%%%%%%%%%%%%%%%%%%%%%%%%%%%%%%%%%%%%%%%%%%%%%%%%%%%%%%%%%%%%%%%%%%%
\section{Probabilistic Age Inference in Twitter}
Our age inference method uses ground-truth labels (users who specify their age), which are then generalised to 700m accounts based on the shared interests, which we derive from Following patterns. 

\subsection{Data Collection and Ground-Truth Labels}
\label{sec:prelim}
To extract ground-truth labels we crawl the Twitter graph and download user descriptions. To do this we implemented a distributed Web crawler using Twitter access tokens mined through several consumer apps. To maximize data throughput while remaining within Twitter's rate limits we built an asynchronous data mining system connected to an access token server using Python's Twisted library~\cite{Wysocki2011}.

\begin{figure}[tb]
\centering
\includegraphics[width=0.8\hsize]{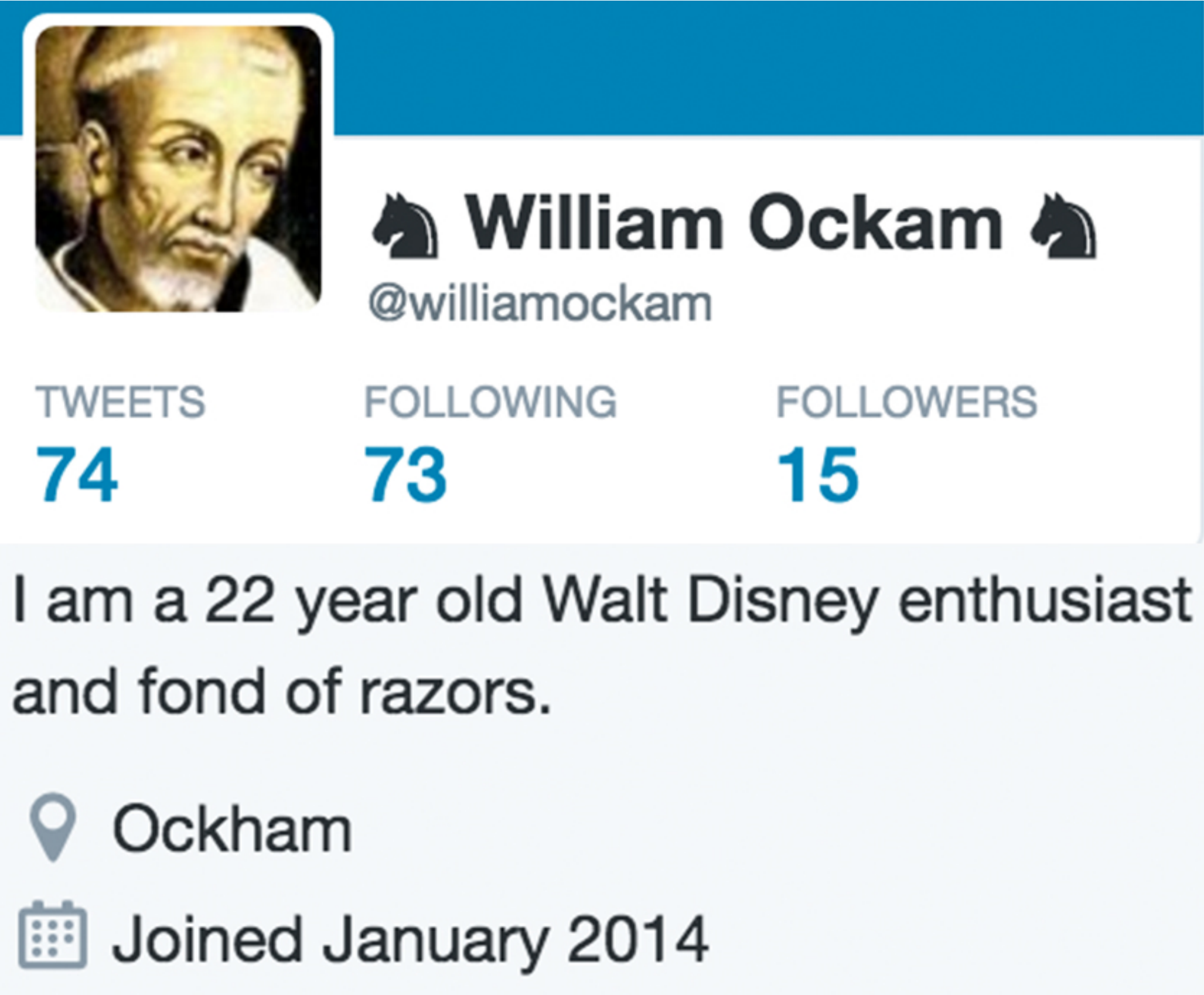}
\caption{A Twitter profile for @williamockam that we created and use to illustrate the method. The profile contains the name, Twitter handle, number of tweets, number of followers, number of people following and a free-text description field with age information.}
\label{fig:twitter_profile}
\end{figure}

% get age information from the data --> ground truth
Our crawl downloaded 700m user description fields. Fig.~\ref{fig:twitter_profile} shows the Twitter profile with associated metadata fields for the fictitious @williamockam account, which we use to illustrate our approach. We index the free-text description fields using Apache SOLR~\citep{Grainger2014} and search the index for REGular EXpression (REGEX) patterns that are indicative of age (e.g., the phrase: ``I am a 22 year old'' in Fig.~\ref{fig:twitter_profile}) across Twitter's four major languages (English, Spanish, French, Portuguese). For repeatability we include our REGEX code in the Appendix.
% Listing~\ref{lst:regex}.\footnote{There are probably more effective ways of generating labels using NLP techniques, but because our main focus is on developing a model for generalising from labels, we leave this for future work.}

%\noindent\begin{minipage}{.35\textwidth}
% \begin{figure*}
% \todo[inline]{I could move this to the supplementary material?}
% \begin{lstlisting}[caption={Regex matching run against Twitter descriptions. The code detects age references in English, German, French and Portuguese. Terms including 'feel like', 'think I am' and 'more / less than' were a major source of error in early versions,  which led us to write a REGEX that explicitly removes them.}, language=bash, basicstyle=\small, showstringspaces=false, label=lst:regex]
% awk '{for (i=2; i<=NF; i+=2) {gsub (/,/, "p1p2p3p4p", $i)} print $0 }' FS="\"" OFS="\"" $x > temp
% awk '{print $2, $3, $6}' FS="," OFS="," temp | 
% sed 's/p1p2p3p4p/\,/g' | egrep -i "[\'a][mn] [][0-9][0-9][ \,\.\!\;y][ \/yea]| 
% [ua]is[ ][0-9][0-9][ \,\.\!\;a][an]|bin[ ][0-9][0-9][ ]| [hg]o[ ][0-9][0-9][ a][an][on]" >> temp1.csv
% sed "s/.*[hghia\'][mnso][ ]\([0-9][0-9]\)[ \,\.\!\;ya]
% [ \/yean].*/\1/I;s/.*bin[ ]\([0-9][0-9]\)[ ].*/\1/I" temp1.csv | 
% egrep -v -i "more than [0-9][0-9]| "think i am [0-9][0-9]|think i'm [0-9][0-9]| 
% i feel like [0-9][0-9] | depuis [0-9][0-9]| [a-ln-su-z] an [0-9][0-9]" > temp2.csv
% awk '{getline a < "temp2.csv"; print $0","a}' temp1.csv > temp3.csv
% \end{lstlisting}
% \end{figure*}
%\end{minipage}
Twitter is ten years old and contains many out-of-date descriptions. To tackle the stale data problem we restricted the ground-truth to active accounts, defined to be accounts that had tweeted or Followed in the last three months (we do not have access to Twitter's logs). This process discovered 133,000 active users who disclosed their age (i.e., $0.02\%$ of the 700m indexed accounts), which we use as ``ground-truth'' labels. For each of these we download every account that they Followed. Fig.~\ref{fig:twitter_profile} shows that @williamockam Follows 73 accounts and we downloaded each of their user IDs.

% \begin{figure}[]
% \centering
% \includegraphics[width=\hsize]{twitter_metadata}
% \caption{The Twitter metadata for @meghatron containing the number of Followers, the number of people she Follows and her number of tweets}
% \label{fig:twitter_metadata}
% \end{figure}

We use ten age categories with a higher resolution in younger ages where there is more labelled data. For our ground-truth data set, the age categories, number of accounts, relative frequency and average number of features per category are shown in Table~\ref{tab:age categories}.
\begin{table}[tb]
  \centering
  \caption{Ground-truth data set: Age categories and counts. ``Mean features'' gives the average number of feature accounts followed.}
    \begin{tabular}{llrrc}
    \toprule
    idx & age range & count & freq & mean features\\
    \midrule
    0     & under 12  & 7,753 & 5.9\% & 23.7\\ 
    1     & 12--13 & 20,851 & 15.8\% & 27.9\\
    2     & 14--15 & 30,570 & 23.1\% & 30.8\\
    3     & 16--17 & 23,982 & 18.1\% & 28.7\\
    4     & 18--24 & 33,331 & 25.2\% & 26.0\\
    5     & 25--34 & 9,286 &  7.0\% & 23.1\\
    6     & 35--44 & 3,046  & 2.3\% & 22.6\\
    7     & 45--54 & 1,838 & 1.0\% & 16.0\\
    8     & 55--64 & 962 & 0.7\% & 11.4\\
    9     & over 65   & 596 & 0.5\% & 11.2\\
    \bottomrule
    \end{tabular}%
  \label{tab:age categories}%
\end{table}% 

% data cleaning
Applying REGEX matches to free-text fields inevitably leads to some false positives due to unanticipated character combinations when working with large data sets. In addition, many Twitter accounts, while correctly labelled, may not represent the interests of human beings. This can occur when accounts are controlled by machines (bots), accounts are set up to look authentic to distribute spam (spam accounts) or account passwords are hacked in order to sell authentic looking Followers. 

% removal of anomalies
To reduce the impact of spurious accounts on the model we note that (1) incorrectly labelled accounts can have a large effect on the model as they are distant in feature space from other members of the class / label (2) incorrectly labelled accounts that have a small effect on the model (eg. because they only follow one popular feature) do not matter much by definition. To measure the effect of each labelled account on the model we compute the Kullback-Leibler divergence $\text{KL}(P||P_{\setminus i})$ between the full model and a model evaluated with one data point missing.
% %
% $
% S_i = \text{KL}(P||P_{\setminus i}),
% $
% %
Here, $P$ is the likelihood of the full, labelled data set, and $P_{\setminus i}$ is the likelihood of the model using the labelled data set minus the $i^{th}$ data point. This methodology identifies any accounts that have a particularly large impact on our predictive distribution. We flagged any training examples that were more than three median absolute deviations from the median score for manual inspection. This process excluded 246 accounts from our training data and examples are shown in Table~\ref{tab:bad_ids}. We also randomly sampled 100 data points from across the full ground-truth set and manually verified them by inspecting the descriptions, tweets and who / what they Follow. 

% Table generated by Excel2LaTeX from sheet 'exclusion_examples'
\begin{table*}[tb]
  \centering
  \caption{Spurious data points identified by taking the Median Absolute Deviation of the leave-one-out KL-Divergence.}
  \resizebox{\textwidth}{!}{%
    \begin{tabular}{rrrr}
    \toprule
    Handle & Twitter Description & REGEX age & Reason to Exclude \\
    \midrule
    RIAMOpera & Opera at the Royal Irish ... Presenting: Ormindo Jan 11... & 11 & An Irish Opera \\
    TiaKeough13 & My name Tia I'm 13 years old. & 13 & Hacked account \\
    39yearoldvirgin & I'm 39 years old... if you're a woman, I want to meet you. & 39    & Probably not 39 \\
    50Plushealths & Retired insurance Agent After 40 years of Services. & retired & Using reciprocation software \\
    MrKRudd & Former PM of Australia... Proud granddad of Josie \& McLean... & grandparent & Outlier. Former AUS PM \\
    \bottomrule
    \end{tabular}%
    }
  \label{tab:bad_ids}%
\end{table*}%

\subsubsection{Public Dataset}

For reproducibility we make an anonymised sample of the data and our code publicly available \footnote{address temporarily removed for anonymity}. The data is in two parts: (1) A sparse bipartite adjacency matrix; (2) a vector of age category labels. This dataset was collected and cleaned according to the methodology described above and then down-sampled to give approximately equal numbers of labels in each of seven classes detailed in Table~\ref{tab:public age categories}. It includes only accounts that explicitly state an age (ie. no grandparents or retirees). The adjacency matrix is in the format of a standard (sparse) design matrix and includes only features that are Followed by at least 10 examples. The high level statistics of this network are described in Table~\ref{tab:public adj stats}.

\begin{table}[tb]
  \centering
  \caption{Public dataset labels: age categories and counts.}
    \begin{tabular}{llrrc}
    \toprule
    idx & age range & count \\
    \midrule
    1     & 10-19 & 4486\\
    2     & 20-29 & 4485\\
    3     & 30-39 & 4487\\
    4     & 40-49 & 4485\\
    5     & 50-59 & 4484\\
    6     & 60-69 & 4481\\
    7     & 70-79 & 4481\\
    \bottomrule
    \end{tabular}%
  \label{tab:public age categories}%
\end{table}% 

\begin{table}[tb]
  \centering
  \caption{Public dataset adjacency matrix statistics. Subscript 1 describes labelled acounts and 2 describes features. V is vertices, E edges and D degree.}
    \begin{tabular}{llrrc}
    \toprule
    attribute & value \\
    \midrule
    $|V_1|$ & 31,389 \\
    $|V_2|$ & 50,190 \\
    $|E|$ & 1,810,569 \\
    avg $D_1$ & 57.7 \\
    max $D_1$ & 2049 \\
    std $D_1$ & 95.2 \\
    avg $D_2$ & 36.1 \\
    max $D_2$ & 4405 \\
    std $D_2$ & 96.2 \\
    \bottomrule
    \end{tabular}%
  \label{tab:public adj stats}%
\end{table}% 

%%%%%%%%%%%%%%%%%%%%%%%%%%%%%%%%%%%%%%%%%%%%%%%%%%%%%%%%
\subsection{Age Inference based on Follows}
\label{sec:method}
% summarize the objective
Given a set of 133,000 labelled data points (ground-truth, i.e., Twitter users who reveal their age) we wish to infer the age of the remaining 700m Twitter users. For this purpose, we define a set of features that can be extracted automatically. The features are based on the Following patterns of Twitter users. Once the features are defined, we propose a scalable probabilistic model for age inference.

\subsubsection{Automatic Feature Selection}
% key hypothesis on which we base our method
Our age inference exploits the hypothesis that someone's interests are indicative of their age, and uses Twitter Follows as a proxy for interests. 
% definition of feature vector
Therefore, the features of our model are the 103,722 Twitter accounts that are Followed by more than ten labelled accounts, which can be found automatically. Of the 73 accounts Followed by @williamockam, 8 had sufficient support to be included in our model. These were: Lord\_Voldemort7, WaltDisneyWorld, Applebees, UniStudios, UniversalORL, HorrorNightsORL, HorrorNights and OlanRogers.

\begin{table*}[tb]
  \centering
  \caption{Follower counts for the eight @williamockam features. The support gives their total number of Followers in our labelled data set and Followers is their total number on Twitter. Fractional counts are from assigning a distribution to grandparents.}
  \resizebox{\textwidth}{!}{%
    \begin{tabular}{l|c|cccccccccc|c}
    \textbf{Twitter Handle} & \textbf{Support} & \textbf{$<$12} & \textbf{12--13} & \textbf{14--15} & \textbf{16--17} & \textbf{18--24} & \textbf{25--34} & \textbf{35--44} & \textbf{45--54} & \textbf{55--64} & \textbf{$\geq$65} & \textbf{Followers} \\
    \hline
    \textbf{Lord\_Voldemort7} & \textbf{273} & 5  & 35  & 75  & 55  & 87  & 13   & 0    & 1    & 1    & 1    & 2.0$\times 10^6$ \\
    \textbf{WaltDisneyWorld} & \textbf{435} & 61  & 100  & 89 & 80  & 65  & 20  & 4 & 7 & 4 & 4 & 2.5$\times 10^6$ \\
    \textbf{Applebees} & \textbf{191} & 18  & 43  & 38  & 30  & 37  & 9   & 8 & 2.33 & 2.33 & 3.33 & 0.57$\times 10^6$ \\
    \textbf{UniStudios} & \textbf{60} & 7  & 7 & 14 & 14 & 13 & 5 & 0 & 0 & 0 & 0 & 0.27$\times 10^6$ \\
     \textbf{UniversalORL} & \textbf{65} & 5  & 13 & 10  & 15 & 14 & 4 & 0 & 1.66 & 1.66 & 0.66 & 0.40$\times 10^6$ 
\\
    \textbf{HorrorNightsORL} & \textbf{5} & 0  & 0  & 0  & 1 & 3 & 1 & 0 & 0 & 0 & 0 & 0.04$\times 10^6$ \\
    \textbf{HorrorNights} & \textbf{18} & 1 & 3 & 1 & 4 & 6 & 0 & 1 & 0.66 & 0.66 & 0.66 & 0.08$\times 10^6$ \\
    \textbf{OlanRogers} & \textbf{16} & 0 & 2 & 0 & 7 & 7 & 0 & 0 & 0 & 0 & 0 & 0.11$\times 10^6$    
    \end{tabular}%
    }
  \label{tab:counts_ockam}%
%  \vspace{-5mm}
\end{table*}%

\begin{table*}[tb]
  \centering
  \caption{Posterior distributions (Equation~\eqref{eq:post0}) for the eight features Followed by @williamockam. Probabilities are $\times 10^{-5}$}
  \resizebox{\textwidth}{!}{%
    \begin{tabular}{l|c|cccccccccc|c}
    \textbf{Twitter Handle} & \textbf{Support} & \textbf{$<$12} & \textbf{12--13} & \textbf{14--15} & \textbf{16--17} & \textbf{18--24} & \textbf{25--34} & \textbf{35--44} & \textbf{45--54} & \textbf{55--64} & \textbf{$\geq$65} & \textbf{Followers} \\
    \hline
    \textbf{Lord\_Voldemort7} & \textbf{273} & 111.7  & 190.9  & 258.0 & 252.3 & 248.6 & 145.9 & 31.9 & 38.9 & 77.6    & 177.5 & 2.0$\times 10^6$ \\
    \textbf{WaltDisneyWorld} & \textbf{435} & 725.0 & 538.2 & 441.2 & 377.6 & 267.3 & 233.2 & 194.2 & 270.7 & 254.5 & 224.4 & 2.5$\times 10^6$ \\
    \textbf{Applebees} & \textbf{191} & 231.8 & 206.3 & 176.6  & 150.3 & 129.8 & 137.4 & 226.7 & 132.4 & 139.6 & 139.2 & 0.57$\times 10^6$ \\
    \textbf{UniStudios} & \textbf{60} & 80.6 & 56.0 & 59.3 & 59.5 & 49.3 & 48.1 & 11.3 & 2.8 & 2.3 & 2.3 & 0.27$\times 10^6$ \\
     \textbf{UniversalORL} & \textbf{65} & 67.4  & 63.0 & 56.6 & 60.5 & 50.7 & 42.0 & 21.1 & 62.7 & 86.4 & 40.6 & 0.40$\times 10^6$ 
\\
    \textbf{HorrorNightsORL} & \textbf{5} & 0.3  & 0.7 & 1.5  & 4.0 & 8.3 & 9.4 & 2.0 & 0.3 & 0.1 & 0.1 & 0.04$\times 10^6$ \\
    \textbf{HorrorNights} & \textbf{18} & 14.0 & 13.7 & 11.3 & 15.5 & 16.1 & 9.4 & 29.1 & 29.9 & 36.8 & 29.3 & 0.08$\times 10^6$ \\
    \textbf{OlanRogers} & \textbf{16} & 4.3 & 9.1 & 10.6 & 21.9 & 19.8 & 5.0 & 1.6 & 1.3 & 1.3 & 1.3 & 0.11$\times 10^6$    
    \end{tabular}%
    }
  \label{tab:posterior_ockam}%
%  \vspace{-5mm}
\end{table*}%

% Example: 5 features 
Table~\ref{tab:counts_ockam} shows the number of labelled accounts Following each @williamockam feature. The support is the number of \textit{labelled} Followers summed over all age categories, while Followers gives the total number of Followers (labelled and unlabelled). 
% add data (grandparents/retirees)
A general trend across all features (not only the ones relevant to @williamockam) is that the age distribution is peaked towards ``younger'' ages as not many older people reveal their age (we show this for the accounts with the highest support in our data set in the Appendix).
To improve the predictive performance of the model in higher age categories we adapted our REGEX to search for grandparents and retirees. This augmented our training data with 176,748 people labelled as retired and 63,895 labelled as grandparents. In our ten-category model, retired people are added to the 65+ category. Grandparents are assigned a uniform distribution across the three oldest age categories, which roughly reflects the age distribution of grandparents in the US~\citep{US2014census}\footnote{
%We acknowledge that the US distribution likely overestimates the ages of the Twitter population. 
This value was used as the US is the largest \emph{Twitter country}.}, such that we ended up with approximately 374,000 labelled accounts in our ground-truth data.

%%%%%%%%%%%%%%%%%%%%%%%%%%%%%%%%%%%%%%%%%%%%%%%%%%%%%%%%%%%%%%
\subsubsection{Probabilistic Model for Age Inference}
 
% Why a Bayesian model
We adopt a Bayesian classification paradigm as this provides a consistent framework to model the many causes of uncertainty (noisy labels, noisy features, survey estimates) encountered in the problem of age inference.

% Some notation We denote the feature vector of an unlabelled account $\chi$ by $X \in \{0,1\}^M$, where $M=103,722$ is the number of features and $X_i = 1$ if and only if the account Follows the $i^{th}$ feature. The corresponding target variable, age, is denoted $A \in\{1,...,10\}$. For @williamockam $A=4$ and $X$ is a vector containing eight ones and 103,714 zeros. We use $\mathbf{X} \in \{0,1\}^{M\times N}$ to represent the features of the whole labelled data set of size $N=374,000$. %\todo[inline]{This should be $\mathbf{A} \in [0,1]^{10 \times N}$, but that might cause notation issues elsewhere}

% objective
Our goal is to predict the age label of an arbitrary Twitter user with feature vector $X$ given the set of feature vectors $\mathbf{X}$ and corresponding ground-truth age labels $\mathbf{A}$. Within a Bayesian framework, we are therefore interested in the posterior predictive distribution 
\begin{align}
P(A | X,\mathbf{X,A}) \propto P(X | A,\mathbf{X,A}) P(A)\,,
\label{eq:prediction}
\end{align}
where $P(A)$ is the prior distribution of Twitter user ages and $P(X | A,\mathbf{X,A})$ the likelihood.

% prior
The prior $P(A)$ is based on a survey of American internet users conducted by~\cite{Duggan2013}. They identified a sample of 1,802 over-18-year olds (speaking either English or Spanish) using random cold calling and recorded their demographic information and use of social media. 288 of their respondents were Twitter users, which  yields a small data set that we can use for the prior prior distributions of over 18s. For under 18s we inferred the corresponding values of the prior using US census data \citep{US2010census}, which leads to our categorical prior
\begin{align}
P(A) = \text{Cat}(\pi) = [1, 2, 2, 3, 14, 23, 23, 22, 6, 4]\times 10^{-2}\,.
\label{eq:prior}
\end{align}

% likelihood
The likelihood $P(X | A,\mathbf{X,A})$ is obtained as follows:
% naive Bayes assumption in likelihood
For scalability we make the Naive Bayes assumption that the decision to Follow an account is independent given the age of the user. This yields the likelihood
\begin{align}
P(X | A,\mathbf{X,A}) = \prod\nolimits_{i=1}^M P(X_i | A,\mathbf{A,X})^{X_i}\,,
\label{eq:likelihood}
\end{align}
where $X_i\in\{0,1\}$ and $i$ indexes the features. $X_i=1$ means ``user $\chi$ Follows feature account $i$''.\footnote{
% explain why we only consider the cases where somebody follows somebody
We only consider cases where $X_i = 1$ since the Twitter graph is sparse: In the full Twitter graph there are $7\times 10^8$ nodes with $5\times 10^{10}$ edges, which implies a density of $1.6\times 10^{-7}$, i.e., the default is to follow nobody. Hence, not following an account does not contain enough information to justify the additional computational cost.}

\begin{figure}[tb]
% \vspace{-9mm}
\centering
    \includegraphics[width=\hsize]{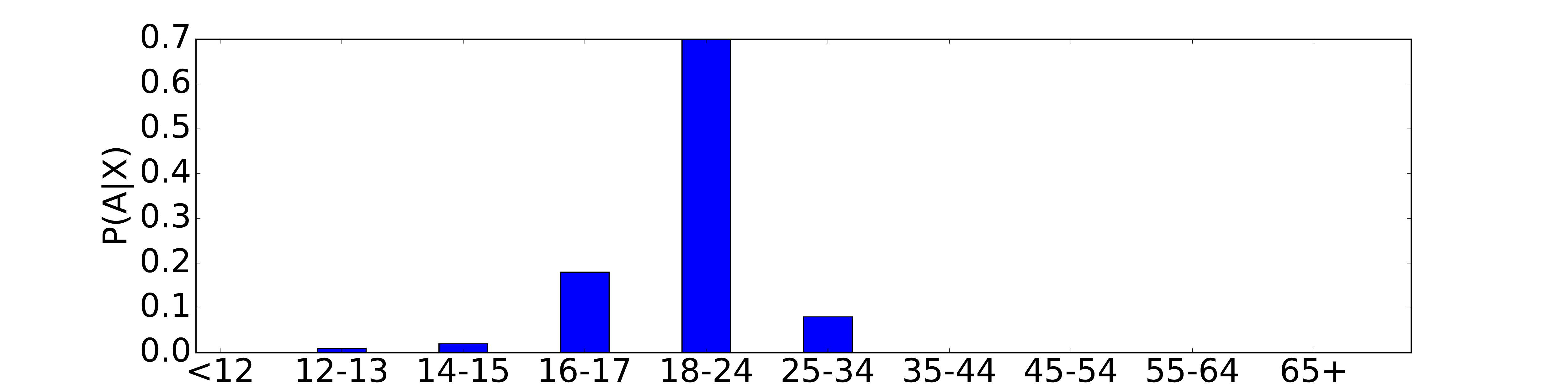}
    \caption{Posterior age distribution for @williamockam.}
\label{fig:williamockam_posterior}
% \vspace{-5mm}
\end{figure}

% Fully Bayesian model for the likelihood
%\todo[inline]{some questions remaining: 1) How does the likelihood depend on the training data? This is very confusing.}
We model the likelihood factors $P(X_i| A,\mathbf{A,X})$ as Bernoulli distributions 
\begin{align}
P(X_i|A = a) = \text{Ber}(\mu_{ia}),
\label{eq:post0}
\end{align}
$i=1,\dotsc,M$,  where $M$ is the number of features and there are 10 age categories indexed by $a =1,\dotsc, 10$. Since our labelled data is severely biased towards ``younger'' age categories we cannot simply learn multinomial distributions $P(A | X_i)$ for each feature based on the relative frequencies of their followers (see Table \ref{tab:age categories}). To smooth out noisy observations of less popular feature accounts we use a hierarchical Bayesian model with conjugate data-dependent Beta priors
\begin{align}
\text{Beta}(\mu_{ia}|b_{ia},c_a)
\end{align}
on the Bernoulli parameters $\mu_{ia}$.
%\todo[inline]{Check the following paragraph again}
% discussion of the choice of hyper-parameters
We seek hyper-parameters $b_{ia},c_{ia}$ of the prior $\text{Beta}(\mu_{ia}|\mathbf{X}, \mathbf{A})$, which do not have a large effect when ample data is available, but produce sensible distributions when it is not. To achieve this we set $c_{a}$ to be constant across all features $X_i$ (hence dropping the $i$ subscript) and proportional to the total number of observations $n_a$ in each age category (the count column in Table~\ref{tab:age categories}). We then set $b_{ia}\propto \tfrac{n_a n_i}{K}$, where $K=7\times 10^8$ is the total number of Twitter users and $n_i$ is the number of Followers of feature $i$ (the Followers column of table~\ref{tab:counts_ockam} for @williamockam's features). Then, the expected prior probability that user $\chi$ Follows account $i$ is 
$
\mathbb{E}[\mu_{ia}|A = a] = \tfrac{b_{ia}}{b_{ia}+c_a}  = \tfrac{n_i}{K+n_i}\,,
$
i.e., it is constant across age classes and varies in proportion to the number of Followers across features. The effect of this procedure is to reduce the model confidence for features where data is limited.
%\todo[inline]{Describe why parameters $c_a$ are shared}
% fully Bayesian model
Due to conjugacy, the posterior distribution on $\mu_{ia}$ is also Beta distributed. Integrating out $\mu_{ia}$ we obtain 
\begin{align} 
P(X_i=1&| A\!=\!a,\mathbf{X,A})
\\
&= 
\int\limits^1_0\! P(X_i\!=\!1|\mu_{i}, A) P(\mu_{i}|\mathbf{X, A},A) d\mu_{i} 
\label{eq:post1}
\\
&=\int\limits^1_0 \mu_{ia} P(\mu_{ia}|\mathbf{X, A}) d\mu_{ia} =\mathbb{E}[\mu_{ia}|\mathbf{X,A}]
\\
&= \tfrac{n_{ia} + b_{ia}}{n_{a}+b_{ia}+c_a} \,,
\label{eq:post2}
\end{align}
where $n_{ia}$ is the number of labelled Twitter users in age category $a$ who Follow feature $X_i$, which are given in Table~\ref{tab:counts_ockam} for the @williamockam features and $n_a$ is the number of Twitter users in category $a$ in the ground-truth (See Table~\ref{tab:age categories}). Performing this calculation yields the likelihoods for the @williamockam features shown in Table~\ref{tab:posterior_ockam}. We are now able to compute the predictive distribution in~\eqref{eq:prediction} to infer the age of an arbitrary Twitter user. The predictive distribution for @williamockam is shown in Figure~\ref{fig:williamockam_posterior} and is calculated by taking the product of the likelihoods from Table~\ref{tab:posterior_ockam} with the prior (Equation~\eqref{eq:prior}) and normalising.

% graphical model
%
% \todo[inline]{May need to change the graphical model such that it only discusses the model, but does not distinguish between training/testing.}
% \begin{figure}[tb]
%   \centering
%     \includegraphics[width=0.5\hsize]{age_pgm3.pdf}
%     \caption{The probabilistic graphical model for the age estimation problem. Circles are random variables, with shading when they are observed. Small dots indicate model parameters and the plates denote replication.}
% \label{fig:pgm}
% \end{figure}
%

The generative process in our model for the likelihood term in~\eqref{eq:prediction} is as follows. 
\begin{enumerate}
\item Draw an age category $A \sim \text{Cat}(\pi)$
\item For each feature $i$ draw $\mu_{ia} \sim \text{Beta}(\mu_{ia}|b_{ia}, c_a)$
\item For each account draw the Follows: $X_i \sim \text{Ber}(\mu_{ia})$
\end{enumerate}

% generative model
%Figure~\ref{fig:pgm} shows the corresponding graphical model that we use to infer the age of Twitter users. The labelled training data is represented as $\mathbf{A}_{i,j}$ and $\mathbf{X}_{i,j}$ and the target value to be predicted as $A$, respectively.  Given the model parameters $\mu_{i,a}$, $A$ only depends on the set of features $X_i$. 

\begin{table*}[tbp]

  \centering
  \caption{The most discriminative features based on the posterior distribution over age (Equation~\eqref{eq:post1}). Descriptions are taken from the $1^{st}$ line of their Wikipedia pages. See the Appendix for a full table with probabilities and handles.}
  \resizebox{\textwidth}{!}{%
  %\scalebox{0.9}{
    \begin{tabular}{rrrrrrrrrr}
 \textbf{$<$12} & \textbf{12--13} & \textbf{14--15} & \textbf{16--17} & \textbf{18--24} & \textbf{25--34} & \textbf{35--44} & \textbf{45--64}\footnote{both categories have the same features} & \textbf{65+} \\
    \hline
    vlogger & child presenter & child singer & singer & metalcore band & hip hop duo & hip hop artist & evangelist & political journalist \\
    minecraft gamer & YouTuber & child singer & metalcore band & rock band & boy band & rapper & evangelist & retired cyclist \\
    internet personality & child actress & child singer & deathcore singer & rapper & boy band & history channel & evangelist & golf channel \\
    vlogger & child actress & child singer & electronic band & computer game & comedian & record label & faith group & retired rugby player \\
    gaming commentator & girl band & child singer & electronic band & rock band & adult actress & boxer & faith magazine & boxer \\
    
   \end{tabular}}
   % }
  \label{tab:all age discriminant feature table}%
\end{table*}%

In Table~\ref{tab:all age discriminant feature table}, we report the five features with the highest posterior age values of $P(A  | X_i = 1)$ for each age category. The account descriptions are taken from the first line of the relevant Wikipedia page. The youngest Twitter users are characterised by an interest in internet celebrities and computer games players. Music genres are important in differentiating all age groups from 12--45. 25--34 year olds are in part marked by entities that saw greater prominence in the past. This group is also distinguished by an interest in pornographic actors. Age categories 45--54 and 55--64 have the same top five and are differentiated by their interest in religious topics. Users older than 65 are identifiable through an interest in certain sports and politics.

%%%%%%%%%%%%%%%%%%%%%%%%%%%%%%%%%%%%%%%%%%%%%%%%%%%%
\section{Experimental Evaluation}
\label{sec:experiments}
% what are we showing?
We demonstrate the viability of our model for age inference in huge social networks by applying it to 700m Twitter accounts. We conducted three experiments: (1) We compare our approach with the language-based model by ~\citet{nguyen2013old}, which can be considered the state of the art for age inference. (2) We compare our age inference results with the survey by \cite{Duggan2013}. (3) We assess the quality of our age inference on a 10\% hold-out set of ground-truth labels and compare it with results obtained from inference based solely on the prior derived from census and survey data in Equation~\eqref{eq:prior} for age prediction. 

\subsection{Comparison with Dutch Language Model}
% comparison with previous results
For comparison with the state-of-the-art work of \cite{nguyen2013old} based on linguistic features (Dutch tweets) we consider the performance of our model as a three-class classifier using the following age bands: under 18, 18--44 and 45+. 

Fig.~\ref{tab:3_class_stats} lists the performance of our age inference algorithm on a 10\% hold-out test set and the Dutch Language Model (DLM) proposed by~\cite{nguyen2013old}. The corresponding performance statistics are shown in Table~\ref{tab:3_class_stats}. 
% Table generated by Excel2LaTeX from sheet 'precision_recall_F120160110-133'
\begin{table*}[t]
  \centering
  \caption{Statistics for age prediction on a held-out test set.}
    \scalebox{1}{
    \begin{tabular}{l|l|rrrr|rrr|rrr}
    & & \textbf{$<$12} &  \textbf{12--13} & \textbf{14--15} & \textbf{16--17} & \textbf{18--24} & \textbf{25--34} & \textbf{35--44} & \textbf{45--54} & \textbf{55--64} & \textbf{$\geq$65} \\
    \hline
   & Test Cases & 651   & 1,731  & 2,678  & 2,036  & 2,670  & 776   & 230   & 5,058  & 5,145  & 20,487 \\
    \hline
\parbox[t]{2mm}{\multirow{3}{*}{\rotatebox[origin=c]{90}{Ours}}} &  Recall & \blue{0.19} & \blue{0.20} & \blue{0.38} & \blue{0.23} & \blue{0.33}  & \blue{0.25} & 0.18 & \blue{0.32} & \blue{0.41} & \blue{0.30} \\
  &  Precision & \blue{0.22} & \blue{0.33} & \blue{0.36} & \blue{0.24} & \blue{0.31} & \blue{0.15} & \blue{0.07} & \blue{0.14} & \blue{0.19} & \blue{0.79} \\
%    \textbf{F1} & 0.21 & 0.27 & 0.37 & 0.24 & 0.32 & 0.19 & 0.10 & 0.20 & 0.26 & 0.43 \\
  &  Micro F1& \multicolumn{4}{l|}{\blue{0.31}} & \multicolumn{3}{l|}{} &\multicolumn{3}{l}{}
%    \textbf{macro F1} & 0.25 &       &       &       &       &       &       &       &       & 
\\
\hline
\parbox[t]{2mm}{\multirow{3}{*}{\rotatebox[origin=c]{90}{S\&C}}} &Recall & 0.01  &  0.02  &  0.02 &  0.03 &  0.14 &   0.23 &  \blue{0.23} &    0.22 &   0.06 &   0.04\\
&Precision & 0.02 &  0.04  &  0.06 &   0.05 &   0.06  &  0.02 &   0.01 &    0.12  &  0.12  &  0.49\\
&Micro F1 & \multicolumn{4}{l|}{0.07} & \multicolumn{3}{l|}{} &\multicolumn{3}{l}{}
    \end{tabular}%
    }
  \label{tab:all_stats}%
\end{table*}%

\begin{table}[tb]
  \centering
  \caption{Performance for three-class age model.}
  \scalebox{0.9}{
\begin{tabular}{l|ccc|ccc}
     & \multicolumn{3}{|c|}{Our Approach} & \multicolumn{3}{c}{DLM~\citep{nguyen2013old}}\\
     & \textbf{$< $18} & \textbf{18--44} & \textbf{$\geq$45} & \textbf{$\leq $18} & \textbf{18--44} & \textbf{$\geq$45} \\
    \hline
    \textbf{Support} & 7,096  & 3,676  & 30,690 & 1,576 & 608 & 310\\
    \textbf{Precision} & 0.76 & 0.39 & \blue{0.96} & \blue{0.93} & \blue{0.67} & 0.82\\
    \textbf{Recall} & 0.68 & 0.50 & \blue{0.95} & \blue{0.98} & \blue{0.75} & 0.45\\
   % \textbf{F1} & 0.72 & 0.44 & 0.96 & N/A & N/A & N/A\\
\hline
    \textbf{Micro F1} & \multicolumn{3}{l|}{\blue{0.86}} &  \multicolumn{3}{l}{\blue{0.86}}%\\
%    \textbf{Macro F1} & 0.58 &    &   &  \blue{0.77}
\end{tabular}%
}
\label{tab:3_class_stats}%
\end{table}
Both methods perform equally well with a Micro F1 score of 0.86. The precision and recall show that the DLM approach is efficient, extracting information from only a small training set (support). This is because significant engineering work went into labelling and feature design. In contrast, our feature generation process is automatic and scalable. While we do not achieve the same performance for the lower age categories, for the oldest age category, our approach performs substantially better than the method by~\cite{nguyen2013old}, suggesting that a hybrid method could perform well. We leave this for future work. 

The major advantages of our model to the state-of-the-art approach are twofold: First, we have applied our age inference to 700m Twitter users, as opposed to being limited to a sample of Dutch Twitter users with a relatively high number of Tweets. Second, generating our training set is fully automatic and relies only on Twitter data\footnote{\cite{nguyen2013old} used additional Linkedin data for labelling}, i.e., no manual labelling or verification is required. 

\begin{figure}[tb]
	\centering
    	\includegraphics[width=\hsize]{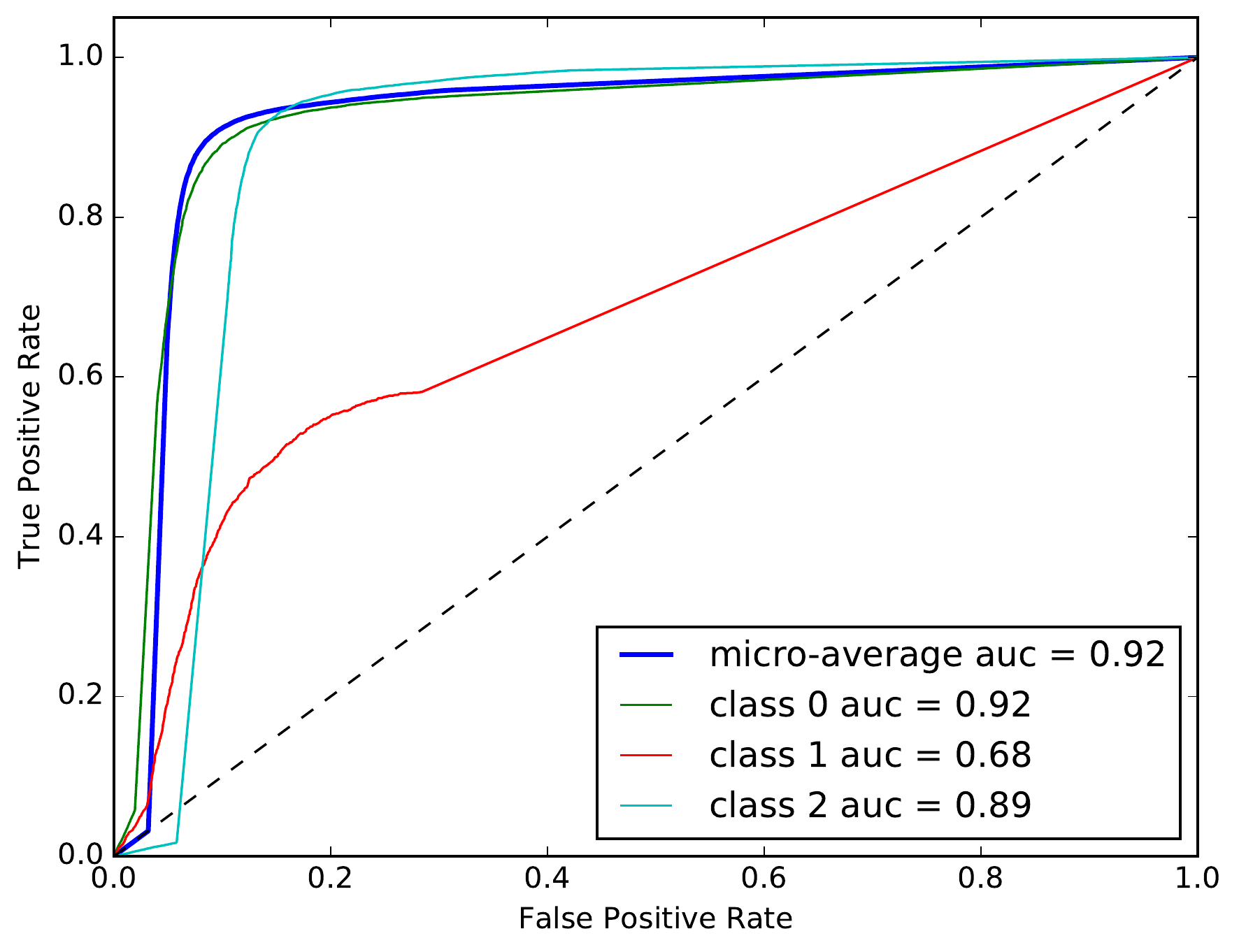}
        \caption{Receiver operator characteristics for three class age detection (0 = under 18, 1 = 18--45, 2 = 45+). The dashed line indicates random performance.}
\label{fig:multiclass_roc_3_class}
\end{figure}
% \begin{figure}[tb]
% 	\centering
%     	\includegraphics[width=\hsize]{1.pdf}
%         \caption{Learning curves showing the classification accuracy on the training and test set as the size of the training set is increased for a three class model. Error bars are one standard deviation on 5-fold cross-validation}
% \label{fig:lc_3_class}
% \end{figure}
%
Fig.~\ref{fig:multiclass_roc_3_class} shows the areas under the receiver-operator characteristics (ROC) curves for our three-class model. The curves are generated by measuring the true positive and false positive rates for each class over a range of classification thresholds. A perfect classifier has an area under the curve (AUC) equal to one, while a completely random classifier follows the dashed line with an $AUC = 0.5$. Performance is excellent for classes under 18 and over 45, but weaker for 18--45 where training data was limited, which we note as an area for improvement in future work. 

\subsection{Comparison with Survey and Census Data}

\begin{figure}[tb]
% \vspace{-9mm}
\centering
    \includegraphics[width=\hsize]{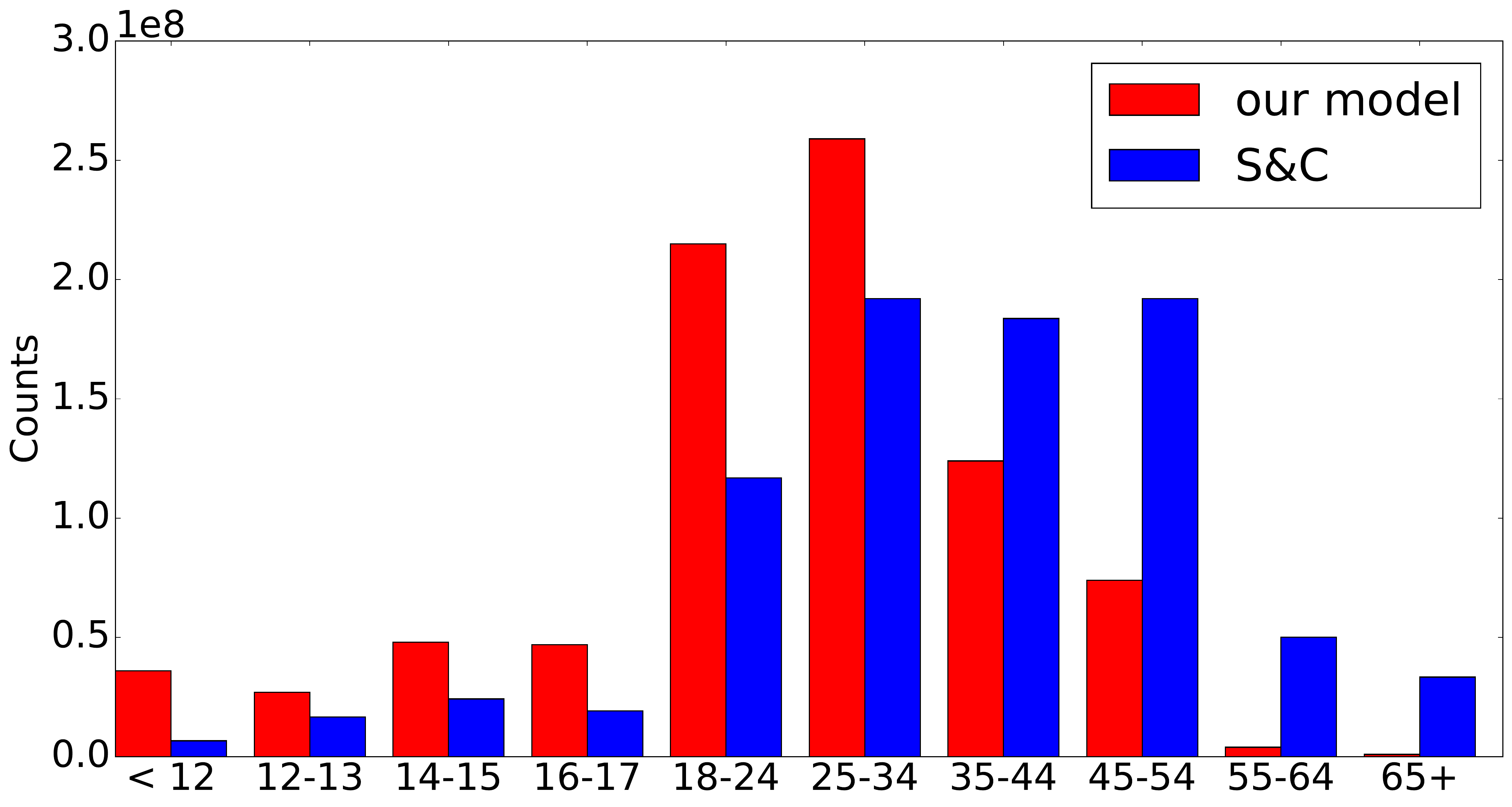}
    \caption{
    Red bars show \#accounts from the 700m unlabelled accounts that our model allocated to each age class using the mode of the predictive posterior. Blue bars show \#accounts that would have been allocated to each age class if  ages were drawn from the Survey and Census (S\&C) prior.}
\label{fig:result_bar}
% \vspace{-5mm}
\end{figure}

We report results on inferring the age of arbitrary Twitter users with the ten category model. Fig.~\ref{fig:result_bar} shows  aggregate classification results for 700m Twitter accounts compared with expected counts based on survey data (S\&C)~\cite{Duggan2013}. Our model predicts that over 50\% of Twitter users are between 18 and 35, i.e., the bias of the original training set has been removed due to the Bayesian treatment. It is likely that S\&C under-represents young people as we did not factor in the increased rates of technology uptake amongst the younger people when converting census data.

\subsection{Quality Assessment}
In the following, we assess the quality of our age inference model (10 categories) on a 10\% hold-out test data set. 

Table~\ref{tab:all_stats} shows the performance statistics for this experiment. The majority of the test cases are in the younger age categories (due to the bias of young people revealing their age) and in older age categories (due to the inclusion of grandparents and retirees). Table~\ref{tab:all_stats} shows that the precision depends on the size of the data (e.g., predicting 25--44 year categories is hard) whereas the recall is fairly stable across all age categories.\footnote{Without the inclusion of grandparents and retirees in the training set, the predictive performance would rapidly drop off for ages greater than 35.} Our model significantly outperforms an approach, which would only be based on the survey and census  data (S\&C), which we use as a prior. This highlights the ability of our model to adapt to the data it actually sees.

%\FloatBarrier

\section{Conclusion}
We proposed a probabilistic model for age inference in Twitter. The model exploits generic properties of Twitter users, e.g., whom/what they follow, which is indicative of their interests and, therefore, their age. Our model performs as well as the current state of the art for inferring the age of Twitter users without being limited to specific linguistic or engineered features. We have successfully applied our model to infer the age of 700 million Twitter users demonstrating the scalability of our approach. 

\appendix

\section{Appendix}
\subsection{Age Extraction Using REGEX Matching of Descriptions}

We extracted user ages from the free text Twitter description using UNIX scripting REGEX matching tools. The exact REGEX strings are included in Listing~\ref{lst:regex}. An initial run of the REGEX revealed some frequent false positives with terms like 'I feel like I am 80' or 'I am more than 10', which were manually corrected for in the final iteration. 

\begin{figure*}
\label{fig:regex}
\begin{lstlisting}[caption={Regex matching run against Twitter descriptions. The code detects age references in English, German, French and Portuguese. Terms including 'feel like', 'think I am' and 'more / less than' were a major source of error in early versions,  which led us to write a REGEX that explicitly removes them.}, language=bash, basicstyle=\small, showstringspaces=false, label=lst:regex]
awk '{for (i=2; i<=NF; i+=2) {gsub (/,/, "p1p2p3p4p", $i)} print $0 }' FS="\"" OFS="\"" $x > temp
awk '{print $2, $3, $6}' FS="," OFS="," temp | 
sed 's/p1p2p3p4p/\,/g' | egrep -i "[\'a][mn] [][0-9][0-9][ \,\.\!\;y][ \/yea]| 
[ua]is[ ][0-9][0-9][ \,\.\!\;a][an]|bin[ ][0-9][0-9][ ]| [hg]o[ ][0-9][0-9][ a][an][on]" >> temp1.csv
sed "s/.*[hghia\'][mnso][ ]\([0-9][0-9]\)[ \,\.\!\;ya]
[ \/yean].*/\1/I;s/.*bin[ ]\([0-9][0-9]\)[ ].*/\1/I" temp1.csv | 
egrep -v -i "more than [0-9][0-9]| "think i am [0-9][0-9]|think i'm [0-9][0-9]| 
i feel like [0-9][0-9] | depuis [0-9][0-9]| [a-ln-su-z] an [0-9][0-9]" > temp2.csv
awk '{getline a < "temp2.csv"; print $0","a}' temp1.csv > temp3.csv
\end{lstlisting}
\end{figure*}

\subsection{The Most Popular Accounts Followed by Labelled Users}

We split the Followers into ten age categories. Table~\ref{tab:counts} shows that general trends across features are that the age distribution is peaked towards ``younger'' ages and that not many older people reveal their age for the top features. 
The Followers column gives the total number of Followers of each feature across the Twitter network. There is a Pearson correlation of 0.86 between the support and the total Follower count for our data set.

\begin{table*}[tb]
  \centering
  \caption{The accounts with the highest support within the labelled data set.}
  \scalebox{0.92}{
    \begin{tabular}{l|c|cccccccccc|c}
    \textbf{Twitter Handle} & \textbf{Support} & \textbf{$<$12} & \textbf{12--13} & \textbf{14--15} & \textbf{16--17} & \textbf{18--24} & \textbf{25--34} & \textbf{35--44} & \textbf{45--54} & \textbf{55--64} & \textbf{$\geq$65} & \textbf{Followers} \\
    \hline
    \textbf{justinbieber} & \textbf{20,359} & 1517  & 5179  & 5737  & 4202  & 3073  & 412   & 99    & 67    & 34    & 38    & 8.7$\times 10^7$ \\
    \textbf{katyperry} & \textbf{18,395} & 1467  & 4180  & 4410  & 3604  & 3575  & 701   & 158   & 124   & 75    & 102   & 9.2$\times 10^7$ \\
    \textbf{taylorswift13} & \textbf{15,199} & 1207  & 3417  & 3674  & 3045  & 2919  & 507   & 113   & 117   & 79    & 122   & 8.1$\times 10^7$ \\
    \textbf{selenagomez} & \textbf{14,264} & 1270  & 3578  & 3691  & 2847  & 2339  & 367   & 76    & 43    & 26    & 27    & 4.6$\times 10^7$ \\
     \textbf{ArianaGrande} & \textbf{13,512} & 1254  & 3404  & 3604  & 2631  & 2172  & 319   & 50    & 40    & 19    & 20    & 4.1$\times 10^7$ 
\\
    \textbf{ddlovato} & \textbf{13,259} & 1099  & 3284  & 3562  & 2741  & 2135  & 301   & 53    & 37    & 19    & 28    & 3.8$\times 10^7$ \\
    \textbf{onedirection} & \textbf{12,834} & 979   & 3472  & 3778  & 2767  & 1622  & 138   & 43    & 20    & 7     & 8     & 3.0$\times 10^7$ \\
    \textbf{Harry\_Styles} & \textbf{12,830} & 912   & 3468  & 3936  & 2751  & 1581  & 120   & 24    & 15    & 9     & 13    & 2.9$\times 10^7$ \\
    \textbf{NiallOfficial} & \textbf{12,498} & 858   & 3431  & 3895  & 2702  & 1468  & 90    & 24    & 15    & 8     & 8     & 2.7$\times 10^7$ \\
    \textbf{YouTube} & \textbf{11,688} & 926   & 2496  & 2687  & 2193  & 2287  & 495   & 183   & 154   & 99    & 169   & 6.4$\times 10^7$    
    \end{tabular}%
    }
  \label{tab:counts}%
%  \vspace{-5mm}
\end{table*}%
% Example: 5 features 

\subsection{The Most Discriminative Features in Each Category}

For each feature we calculate the posterior probability of Following that feature given the user's age. We sort the posteriors within each age category and present the accounts with the five highest values in Table~\ref{tab:full age discriminant feature table}.

\begin{table*}[tbp]
  \centering
  \caption{In the model the features are popular Twitter accounts. This table contains the posterior distributions $p(X=1 | A=a)$ over age for the five most discriminative (useful) features in each age class.}
  \scalebox{0.9}{
    \begin{tabular}{llrrrrrrrrrr}
    \textbf{twitter\_handle} & \textbf{description} & \textbf{$<$12} & \textbf{12--13} & \textbf{14--15} & \textbf{16--17} & \textbf{18--24} & \textbf{25--34} & \textbf{35--44} & \textbf{45--54} & \textbf{55--64} & \textbf{65+} \\
    \hline
    \multicolumn{6}{l}{\blue{Under 12-year olds}}\\
    \textbf{RosannaPansino} & vlogger & \blue{0.40}  & 0.22  & 0.15  & 0.09  & 0.07  & 0.02  & 0.01  & 0.01  & 0.01  & 0.02 \\
    \textbf{AntVenom} & minecraft gamer & \blue{0.40}  & 0.25  & 0.15  & 0.09  & 0.06  & 0.02  & 0.01  & 0.01  & 0.01  & 0.01 \\
    \textbf{Bajan\_Canadian} & internet personality & \blue{0.37}  & 0.25  & 0.17  & 0.10  & 0.06  & 0.02  & 0.00  & 0.01  & 0.01  & 0.01 \\
    \textbf{shaycarl} & vlogger & \blue{0.36}  & 0.20  & 0.14  & 0.10  & 0.07  & 0.04  & 0.02  & 0.02  & 0.02  & 0.02 \\
    \textbf{InTheLittleWood} & gaming commentator & \blue{0.34}  & 0.23  & 0.16  & 0.11  & 0.08  & 0.02  & 0.01  & 0.01  & 0.01  & 0.02 \\
    \hline
   \multicolumn{6}{l}{\blue{12--13 year olds}}\\
    \textbf{ivandorschner} & child TV presenter & 0.18  & \blue{0.27}  & 0.20  & 0.11  & 0.09  & 0.03  & 0.03  & 0.02  & 0.03  & 0.04 \\
    \textbf{Vikkstar123} & youtuber & 0.29  & \blue{0.26}  & 0.20  & 0.14  & 0.07  & 0.02  & 0.01  & 0.01  & 0.01  & 0.02 \\
    \textbf{PeytonList} & child actress & 0.29  & \blue{0.25}  & 0.20  & 0.14  & 0.07  & 0.02  & 0.01  & 0.01  & 0.01  & 0.01 \\
    \textbf{G\_Hannelius} & child actress & 0.31  & \blue{0.25}  & 0.18  & 0.13  & 0.07  & 0.02  & 0.02  & 0.01  & 0.01  & 0.01 \\
    \textbf{Cimorelliband} & girlband & 0.20  & \blue{0.25}  & 0.23  & 0.17  & 0.09  & 0.02  & 0.01  & 0.01  & 0.01  & 0.01 \\
    \hline
    \multicolumn{6}{l}{\blue{14--15 year olds}}
    \\
    \textbf{therealsavannah} & child pop singer & 0.10  & 0.18  & \blue{0.27}  & 0.21  & 0.12  & 0.02  & 0.01  & 0.03  & 0.03  & 0.03 \\
    \textbf{jessicajarrell} & child pop singer & 0.12  & 0.21  & \blue{0.26}  & 0.24  & 0.10  & 0.02  & 0.01  & 0.01  & 0.01  & 0.01 \\
    \textbf{TheDylanHolland} & child R\&B singer & 0.12  & 0.22  & \blue{0.26}  & 0.24  & 0.11  & 0.02  & 0.01  & 0.01  & 0.01  & 0.01 \\
    \textbf{OfficialBirdy} & child singer & 0.10  & 0.17  & \blue{0.26}  & 0.24  & 0.13  & 0.04  & 0.01  & 0.02  & 0.02  & 0.02 \\
    \textbf{officialjman} & child singer & 0.10  & 0.18  & \blue{0.26}  & 0.28  & 0.13  & 0.02  & 0.01  & 0.01  & 0.01  & 0.01 \\
    \hline
    \multicolumn{6}{l}{\blue{16--17 year olds}}\\
    \textbf{TannerPatrick} & singer & 0.05  & 0.13  & 0.25  & \blue{0.30}  & 0.18  & 0.03  & 0.01  & 0.01  & 0.01  & 0.01 \\
    \textbf{TheWordAlive} & metalcore band & 0.04  & 0.11  & 0.19  & \blue{0.29}  & 0.22  & 0.09  & 0.02  & 0.01  & 0.01  & 0.01 \\
    \textbf{MitchLuckerSS} & deathcore singer & 0.05  & 0.14  & 0.23  &\blue{0.29}  & 0.20  & 0.04  & 0.01  & 0.01  & 0.01  & 0.02 \\
    \textbf{metrostation} & electronic band & 0.03  & 0.07  & 0.15  &\blue{0.29}  & 0.18  & 0.10  & 0.04  & 0.06  & 0.06  & 0.03 \\
    \textbf{BreatheCarolina} & electronic band & 0.06  & 0.15  & 0.22  & \blue{0.29}  & 0.19  & 0.06  & 0.01  & 0.01  & 0.01  & 0.01 \\
    \hline
    \multicolumn{6}{l}{\blue{18--24 year olds}}\\
    \textbf{wecameasromans} & metalcore band & 0.05  & 0.13  & 0.22  & 0.28  & \blue{0.21}  & 0.06  & 0.01  & 0.01  & 0.01  & 0.01 \\
    \textbf{Sum41} & rock band & 0.07  & 0.11  & 0.18  & 0.24  &\blue{0.21}  & 0.09  & 0.02  & 0.02  & 0.03  & 0.03 \\
    \textbf{hopsin} & rapper & 0.04  & 0.09  & 0.13  & 0.19  & \blue{0.20}  & 0.09  & 0.09  & 0.06  & 0.05  & 0.07 \\
    \textbf{Diablo} & computer game & 0.03  & 0.06  & 0.09  & 0.13  & \blue{0.20}  & 0.17  & 0.09  & 0.05  & 0.06  & 0.12 \\
    \textbf{paparoach} & rock band & 0.04  & 0.09  & 0.14  & 0.19  & \blue{0.20}  & 0.12  & 0.07  & 0.06  & 0.06  & 0.04 \\
    \hline
    \multicolumn{6}{l}{\blue{25--34 year olds}}\\
    \textbf{icp} & hip hop duo & 0.02  & 0.04  & 0.05  & 0.09  & 0.19  & \blue{0.37}  & 0.09  & 0.04  & 0.05  & 0.05 \\
    \textbf{kevinrichardson} & boyband member & 0.02  & 0.03  & 0.05  & 0.09  & 0.16  & \blue{0.35}  & 0.12  & 0.07  & 0.06  & 0.04 \\
    \textbf{skulleeroz} & boyband member & 0.02  & 0.04  & 0.06  & 0.09  & 0.16  & \blue{0.33}  & 0.12  & 0.07  & 0.06  & 0.05 \\
    \textbf{LeeEvansNews} & comedien & 0.02  & 0.03  & 0.06  & 0.07  & 0.17  & \blue{0.32}  & 0.09  & 0.08  & 0.09  & 0.09 \\
    \textbf{miko\_lee} & adult actress & 0.04  & 0.03  & 0.03  & 0.05  & 0.17  & \blue{0.31}  & 0.08  & 0.07  & 0.08  & 0.14 \\
    \hline
    \multicolumn{6}{l}{\blue{35--44 year olds}}\\
    \textbf{djspooky} & hip hop artist & 0.01  & 0.02  & 0.03  & 0.02  & 0.04  & 0.15  & \blue{0.45}  & 0.14  & 0.06  & 0.08 \\
    \textbf{Mr\_Mike\_Jones} & rapper & 0.01  & 0.01  & 0.01  & 0.01  & 0.03  & 0.14  & \blue{0.44}  & 0.16  & 0.09  & 0.10 \\
    \textbf{HISTORYTV18} & history  TV channel & 0.02  & 0.03  & 0.03  & 0.05  & 0.09  & 0.14  & \blue{0.36}  & 0.10  & 0.06  & 0.13 \\
    \textbf{TopDawgEnt} & record label & 0.03  & 0.07  & 0.05  & 0.07  & 0.11  & 0.11  & \blue{0.36}  & 0.09  & 0.03  & 0.07 \\
    \textbf{DannySwift} & boxer & 0.02  & 0.03  & 0.04  & 0.07  & 0.06  & 0.09  & \blue{0.33}  & 0.12  & 0.08  & 0.16 \\
    \hline
    \multicolumn{10}{l}{\blue{45--54 and 55--64-year olds} (identical most-discriminant features)}\\
    \textbf{JohnBevere} & evangelist & 0.00  & 0.00  & 0.00  & 0.00  & 0.01  & 0.01  & 0.07  & \blue{0.36}  & \blue{0.39}  & 0.15 \\
    \textbf{edstetzer} & evangelist & 0.00  & 0.00  & 0.00  & 0.00  & 0.00  & 0.01  & 0.07  & \blue{0.36}  & \blue{0.39}  & 0.16 \\
    \textbf{ChristineCaine} & evangelist & 0.00  & 0.00  & 0.01  & 0.00  & 0.01  & 0.01  & 0.07  & \blue{0.36}  & \blue{0.38}  & 0.15 \\
    \textbf{womenoffaith} & faith group & 0.00  & 0.00  & 0.00  & 0.00  & 0.00  & 0.02  & 0.08  & \blue{0.36}  & \blue{0.38}  & 0.16 \\
    \textbf{RELEVANT} & faith magazine & 0.00  & 0.01  & 0.00  & 0.01  & 0.01  & 0.01  & 0.07  & \blue{0.35}  & \blue{0.38}  & 0.17 \\
    \hline
    \multicolumn{6}{l}{\blue{People over 65}}\\
    \textbf{afneil} & political journalist & 0.00  & 0.00  & 0.01  & 0.01  & 0.02  & 0.02  & 0.04  & 0.17  & 0.25  & \blue{0.48} \\
    \textbf{Chris\_Boardman} & retired cyclist & 0.01  & 0.01  & 0.01  & 0.02  & 0.01  & 0.01  & 0.04  & 0.17  & 0.25  & \blue{0.47} \\
    \textbf{SkySportsGolf} & golf TV channel & 0.01  & 0.02  & 0.02  & 0.02  & 0.03  & 0.01  & 0.04  & 0.16  & 0.22  & \blue{0.46} \\
    \textbf{IamAustinHealey} & retired rugby player & 0.04  & 0.02  & 0.01  & 0.01  & 0.01  & 0.01  & 0.04  & 0.17  & 0.25  & \blue{0.45} \\
    \textbf{anthonyfjoshua} & boxer & 0.02  & 0.03  & 0.03  & 0.04  & 0.09  & 0.06  & 0.03  & 0.08  & 0.15  & \blue{0.45}
    \end{tabular}%
    }
  \label{tab:full age discriminant feature table}%
\end{table*}%

% % failure of the point estimate
% After applying Laplacian smoothing with parameter equal to 1~\cite{Bishop2006} to $P(X_i = 1 | A,\mathbf{A,X})$ the point estimate model\todo{which equation?} produces many pathological distributions with high probability in incorrect categories or bi-modal class predictions of the form ``under 12 with 50\% probability and over 65 with 50\% probability''. The weakness of this model is that it assumes the same level of uncertainty for each $P(X_i = 1 | A,\mathbf{A,X})$ when in reality the distributions for Twitter's most popular accounts are less uncertain than those with only a few thousand followers.

%\FloatBarrier

% \section{Acknowledgements}
% This work was partly funded through a Royal Commission for the Exhibition of 1851 Industrial Fellowship.

% %%%%%%%%%%%%%%%%%%%%%%%%%%%%%%%%%%%%%%%%%%%%%%%%%%%%%%%%%%%%%%

\bibliographystyle{aaai}

\end{document}